\documentclass{article} 
\usepackage{amsmath,amssymb,booktabs,multirow,graphicx,xfrac}
\begin{document}

%\preprint{}

\title{Effects of mutual coupling in\\dual-resonance metamaterials} 

\date{}
\maketitle 
\begin{center}
Withawat Withayachumnankul$^{1,2}$, Christophe Fumeaux$^{1}$, and Derek Abbott$^{1}$

$^1$School of Electrical \& Electronic Engineering, The University of Adelaide,
Adelaide, SA 5005,
Australia

$^2$School of Electronic Engineering, Faculty of Engineering, King Mongkut's Institute of Technology Ladkrabang, Bangkok 10520, Thailand

\end{center}

\begin{abstract} 
This Letter presents an investigation on the effects of mutual coupling in a metamaterial comprising two sets of electric-LC (ELC) resonators with different resonance frequencies. Through simulation and experiment, it is found that the two resonances experience significant shifting and weakening as they become spectrally close. An equivalent circuit model suggests that inductive coupling among the two resonator sets is a primary cause of the change in the resonance properties. This study is fundamental to designing metamaterials with an extended bandwidth or spatially variable response.
\end{abstract}

\section{Introduction}

A metamaterial defines a group of resonators that collectively exhibit a strong electric and/or magnetic resonance. These resonators operate in the effective medium regime owing to their subwavelength dimensions. Thus, their response can be characterized by a homogenized permittivity and permeability, which can be controlled via the shape, size, and materials of the resonators. In the last decade, metamaterials have appeared in various applications across the electromagnetic spectrum \cite{Wit09}.

To achieve desirable properties, the knowledge on inter-resonator coupling is significant, since this near-field behavior directly influences the effective response of metamaterials. Earlier studies have focused on coupling among identical resonators \cite{Liu08,Bit09,Sin09,Sin10,Fet10}. Recently proposed metamaterials contain different unit cells to overcome the inherent narrow-band limitation \cite{Yua08a,Yua08b,Bin08,Hoo07,Wen09,Tao10,Liu11}. Additionally, transformation optics requires a spatial change in the metamaterial response \cite{Che10,Gol11}. Hence, the study on the interaction of different resonators become vital. 

In this work, mutual coupling in a dual-resonance metamaterial, composed of different electric-LC  (ELC) resonators \cite{Sch06,Pad07,Wit10}, is explored. The coupling effect is observed through simulation and experiment. A microwave metamaterial is used for ease of continuous resonance tuning via integrated varactors \cite{Wit11}. A double-tuned resonance circuit along with a transmission line model provides further insight into the coupling phenomenon. 

\section{Experiment}

\begin{figure}[!b]
\centering\includegraphics{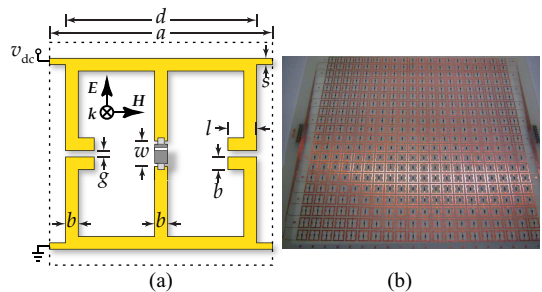}
\caption{(a) ELC resonator with a varactor loaded at the central gap. The dc bias lines are positioned at the four corners. The dotted lines bound a unit cell. (b) Fabricated planar metamaterial containing 20$\times$20 resonators. The biasing strips are visible on the left and right.}
\label{fig:TE4_resonators}
\end{figure}

A unit cell of the varactor-integrated ELC resonator is illustrated in Fig.~\ref{fig:TE4_resonators}(a). The dimensions of the resonator are as follows: $a=13$~mm, $d=11$~mm, $l=1.8$~mm, $w=1.7$~mm, $b=0.8$~mm, $g=0.4$~mm, and $s=0.3$~mm. Each resonator is made of 35-$\mu$m-thick bare copper, supported by an epoxy FR4 substrate with a thickness of 0.8~mm, a dielectric constant of 4.5, and a loss tangent of 0.02. The entire metamaterial panel comprises 20$\times$20 resonators. The varactor is Infineon BB837 with a tunable range from 9.5 pF down to 0.58~pF for the reverse bias from 0 to 24~V. The varactors in each row are connected in parallel to an independent dc source. Fig.~\ref{fig:TE4_resonators}(b) shows the fabricated metamaterial including dc feeds.

The experiment is carried out in an anechoic chamber with a vector network analyzer connecting to transmitting and receiving horn antennas. The sample transmission is measured and normalized by the free-space transmission. The simulation is performed using Ansoft HFSS employing the finite-element method. Periodic boundary conditions are used for the transverse boundaries to replicate an infinite 2D resonator array. The varactor is substituted by a lumped capacitor, whose capacitance is available from the SPICE model. 

\begin{figure}
\centering\includegraphics{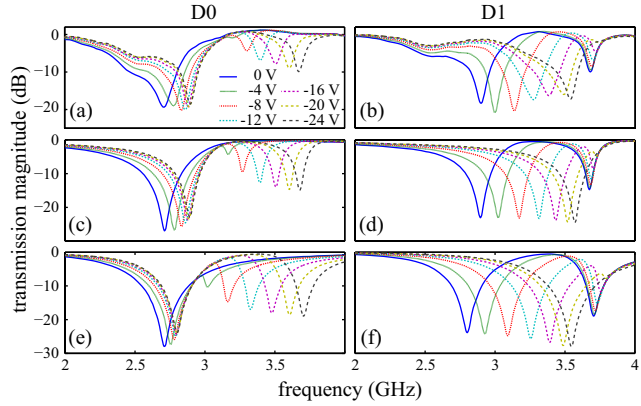}	% MATLAB figure scaled to 25%
\caption{Transmission magnitudes from experiment (a,b), simulation (c,d), and circuit model (e,f). (a,c,e) The odd-row bias is 0~V, whilst the even-row bias varies from 0 to -24~V (case D0). (b,d,f) The odd-row bias is -24~V, whilst the even-row bias varies from 0 to -24~V (case D1).}
\label{fig:TE4_double_res}
\end{figure}

The experiment comprises three cases. First, all of the varactors in the array are connected to a dc bias that changes from 0 to -24 V with an interval of -4 V (case S), the results of which are elaborated in \cite{Wit11}. Next, the varactors in the odd rows are grounded, whilst those in the even rows are connected to a bias varying from 0 to -24 V with a step of -4 V (case D0). Finally, the odd-row varactors are pulled to -24 V, and the even-row bias varies from 0 to -24 V (case D1). The experimental results for cases D0 and D1 are shown in Fig.~\ref{fig:TE4_double_res}(a,b), whilst Fig.~\ref{fig:TE4_double_res}(c,d) are the corresponding simulation results. The simulation and experiment yield similar results with small discrepancies. The transmission magnitudes for both cases D0 and D1 reveal the lower and higher resonances, caused by lower and higher absolute biases, respectively.

\begin{figure}
\centering\includegraphics{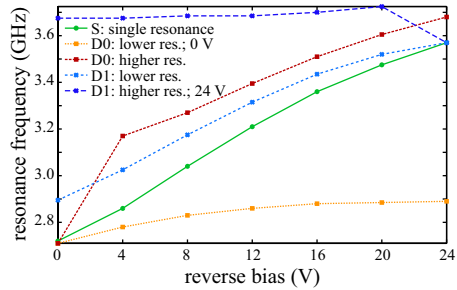} % MATLAB figure scaled to 15%
\caption{Simulated resonance frequencies as a function of bias voltage for three different bias cases: S, D0, and D1.}
\label{fig:TE4_resonance_freq}
\end{figure}

The resonance frequencies as a function of the bias voltage are depicted in Fig.~\ref{fig:TE4_resonance_freq}.
In both cases D0 and D1, the lower and higher resonances experience a slight blueshift, relative to the resonance of the single-biased metamaterial (case S) at the same bias voltage. This blueshift is related to the fact that the inter-cell coupling, typically reinforced among the ELC resonators with the same resonance frequency, is weakened as half of the resonators are spectrally tuned away.

%The resonance frequencies as a function of the bias voltage are depicted in Fig.~\ref{fig:TE4_resonance_freq}. In case D0, the lower resonance experiences a slight blueshift, as the higher resonance shifts to high frequencies, following the increasing reverse bias. When compared to the single-biased metamaterial (case S) at the same bias voltage, case D0 always exhibits both of the resonances at higher frequencies. Since ELC resonators typically couple to their neighbors constructively, tuning half of the resonators towards higher frequencies results in a weaker inter-cell coupling and hence a blueshift in the resonance. This explanation is also applicable to case D1, where both the lower and higher resonances experience a blueshift relative to the resonance of the single-biased metamaterial. 

In both cases D0 and D1, the strength and linewidth of the higher resonance are noticeably smaller than those of the lower counterpart, and the difference is more pronounced when the two resonances come closer to each other. This effect, also observable in \cite{Yua08a,Yua08b}, is attributed to the different oscillation phases in the lower-biased resonators. At the lower resonance frequency, the in-phase oscillation induces constructive current in the higher-biased resonators. Slightly above the lower resonance frequency, the out-of-phase oscillation \cite{Pen04,Smi04} induces destructive current in the higher-biased resonators. Hence, the imbalance in the two resonances is observed.

\section{Equivalent circuit model}

\begin{figure}[!b]
\centering\includegraphics{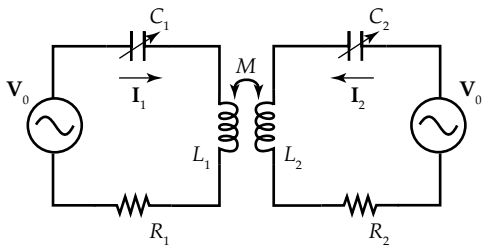}
\caption{Double-tuned resonant circuit. The phasor voltage and current are represented by $\mathbf{V}_0$ and $\mathbf{I}_{\{1,2\}}$, respectively.}
\label{fig:TE4_circuit}
\end{figure}

A double-tuned resonant circuit model shown in Fig.~\ref{fig:TE4_circuit} is utilized to gain further insight on the coupling mechanism of the complex dual-resonance metamaterial \cite{Oha07}. It is composed of two inductively coupled series RLC circuits. This simplification lies on the basis that an ELC resonator can be approximated by a series RLC circuit with a Lorentzian response and that the magnetic flux linkage exists between neighboring resonators. The two resonant circuits in Fig.~\ref{fig:TE4_circuit} share the same ac source, since their equivalent ELC resonators are uniformly excited.

Through Kirchhoff's current law, the system of equations for the circuit in Fig.~\ref{fig:TE4_circuit} is given as
\begin{eqnarray}\label{eq:TE4_kcl}
\begin{bmatrix}
	\mathbf{Z}_1 & \mathbf{Z}_M           \\
	\mathbf{Z}_M & \mathbf{Z}_2 
 \end{bmatrix}
\begin{bmatrix}
	\mathbf{I}_1       \\
	\mathbf{I}_2 
 \end{bmatrix}
=
\begin{bmatrix}
	\mathbf{V}_0       \\
	\mathbf{V}_0 
 \end{bmatrix}\;,
\end{eqnarray}
with complex impedances  $\mathbf{Z}_{\{1,2\}}=R_{\{1,2\}}+j\omega L_{\{1,2\}} +1/j\omega C_{\{1,2\}}$ and $\mathbf{Z}_M=j\omega M$. The sign for the mutual inductance, $M$, depends on the magnetic field activity. It is positive if the overall mutual induction is constructive, and becomes negative otherwise. From Eq.~\ref{eq:TE4_kcl} the impedance seen by the source $\mathbf{V}_0$ can be described as
\begin{equation}
\mathbf{Z}_c = \frac{\mathbf{V}_0}{\mathbf{I}_1+\mathbf{I}_2} = \frac{\mathbf{Z}_1\mathbf{Z}_2-\mathbf{Z}_M^2}{\mathbf{Z}_1+\mathbf{Z}_2-2\mathbf{Z}_M}\;.
\end{equation}
Transmission line theory is used in conjunction with this circuit model to determine the transmission characteristics \cite{Oha07,Aza08}. The substrate introduces a parallel impedance $Z_s$ to $\mathbf{Z}_c$, and also causes a change in the transmission magnitude at its back surface. Hence, the model complex transmission coefficient is given as
\begin{eqnarray}\label{eq:TE4_transmission}
\mathbf{T} = \frac{2\mathbf{Z}_\mathrm{eff} }{\mathbf{Z}_\mathrm{eff} + Z_0} \frac{2Z_0}{Z_s+Z_0};\quad \mathbf{Z}_\mathrm{eff} = \mathbf{Z}_c || Z_s\;,
\end{eqnarray}
where the free-space and substrate intrinsic impedances, $Z_0$ and $Z_s$, equal 377 and 177.72 $\Omega$, respectively.

Eq.~\ref{eq:TE4_transmission} contains seven variables: $R_{\{1,2\}}$, $C_{\{1,2\}}$, $L_{\{1,2\}}$, and $M$. Here, $R_1 = R_2 = R_s$ and $L_1 = L_2 = L_s$, because of an identical resonator structure. The capacitances,  $C_{\{1,2\}}$, equal the variable capacitances  $C_{\{d1,d2\}}$ in series with the structure capacitance $C_s$. Although $C_{\{d1,d2\}}$ are correlated with the varactor capacitances, and $C_s$, and $L_s$ can be readily obtained \cite{Wit11}, using these values do not yield satisfying results. This is because the simplified circuit model represents a large resonator array, and thus the individual capacitors and inductors in the model are different from those in the array. Hence, only a relative change is deduced from these known values. 

Based on these assumptions, the fitted values are as follows: $R_s$ = 12~$\Omega$, $C_s$ = 92~fF, $L_s$ = 36.5~nH, $M$ = 2.8~nH. The variable capacitance $C_d$ = \{1.9, 0.62, 0.32, 0.21, 0.16, 0.13, 0.12\} pF for the reverse bias varying from 0 to 24~V in 4-V steps. It is clear from the transmission results in Fig.~\ref{fig:TE4_double_res}(e,f) that the circuit model can mimic the behavior of the dual-resonance metamaterial. The same parameter set also produces an agreeable transmission magnitude for the single-resonance metamaterial. A small difference in the resonance shape is mainly caused by the high-frequency dipole resonance \cite{Aza08}, which is not taken into account by the model.

\begin{figure}
\centering\includegraphics{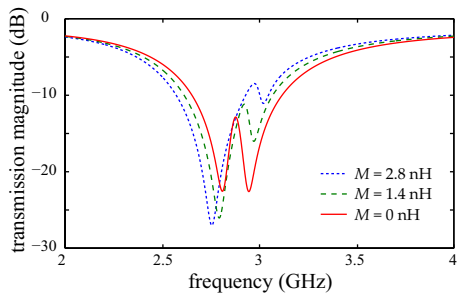} % MATLAB figure scaled to 15%
\caption{Transmission magnitudes for various mutual inductance $M$ obtained from the circuit model. The odd- and even-row biases are fixed to 0 and 4~V, respectivelyใ}
\label{fig:TE4_decoupling}
\end{figure}

The mutual inductance $M$ implies the magnetic flux linkage between the two resonator groups. By changing this value, the influence of magnetic coupling on the metamaterial response can be observed. Fig.~\ref{fig:TE4_decoupling} depicts the model transmission magnitudes following a change in $M$. As $M$ is reduced, the strength of the lower and higher resonances decreases and increases, respectively. Additionally, the lower resonance undergoes a blueshift, whilst the higher resonance undergoes a redshift. This effect can be ascribed to a reduction in the constructive and destructive magnetic coupling among the different resonators at the lower and higher resonance frequencies, respectively. At $M=0$, where the two systems are decoupled, the two resonances become comparable in strength. Hence, a difference in the resonance strengths is responsible purely by the inter-cell inductive coupling.

\section{Conclusion}

In conclusion, the effects of mutual coupling in a dual-resonance metamaterial have been studied through the simulation, experiment, and interpreted through a circuit model. As a result of constructive and destructive interferences in the magnetic activity, the coupled resonances are unbalanced and shifted considerably. This study is fundamental towards designing extended-bandwidth or spatially dependent metamaterials, particularly relevant to the emerging field of transformation optics. The results from this study are applicable to other types of resonators possessing a similar resonance mode in any frequency range. Future work involves minimising the coupling effect via optimal designs.

\section*{Acknowledgement}

The authors thank I. Linke, B. Pullen, P. Simcik, and H. Ho for their technical assistance. This research was supported by the Australian Research Council Discovery Projects funding scheme under Project DP1095151.

%\bibliography{2011_TE4_arXiv}

\begin{thebibliography}{10}

\bibitem{Wit09}
W.~Withayachumnankul and D.~Abbott, ``Metamaterials in the terahertz regime,''
  {\em IEEE Photonics J.}~{\bf 1}(2), pp.~99--118, 2009.

\bibitem{Liu08}
N.~Liu, S.~Kaiser, and H.~Giessen, ``Magnetoinductive and electroinductive
  coupling in plasmonic metamaterial molecules,'' {\em Adv. Mater.}~{\bf
  20}(23), pp.~4521--4525, 2008.

\bibitem{Bit09}
A.~Bitzer, J.~Wallauer, H.~Merbold, H.~Helm, T.~Feurer, and M.~Walther,
  ``Lattice modes mediate radiative coupling in metamaterial arrays,'' {\em
  Opt. Express}~{\bf 17}(24), pp.~22108--22113, 2009.

\bibitem{Sin09}
R.~Singh, C.~Rockstuhl, F.~Lederer, and W.~Zhang, ``The impact of nearest
  neighbor interaction on the resonances in terahertz metamaterials,'' {\em
  Appl. Phys. Lett.}~{\bf 94}, p.~021116, 2009.

\bibitem{Sin10}
R.~Singh, C.~Rockstuhl, and W.~Zhang, ``Strong influence of packing density in
  terahertz metamaterials,'' {\em Appl. Phys. Lett.}~{\bf 97}, p.~241108, 2010.

\bibitem{Fet10}
N.~Feth, M.~K\"onig, M.~Husnik, K.~Stannigel, J.~Niegemann, K.~Busch,
  M.~Wegener, and S.~Linden, ``Electromagnetic interaction of split-ring
  resonators: The role of separation and relative orientation,'' {\em Opt.
  Express}~{\bf 18}(7), pp.~6545--6554, 2010.

\bibitem{Yua08a}
Y.~Yuan, C.~Bingham, T.~Tyler, S.~Palit, T.~H. Hand, W.~J. Padilla, D.~R.
  Smith, N.~M. Jokerst, and S.~A. Cummer, ``Dual-band planar electric
  metamaterial in the terahertz regime,'' {\em Opt. Express}~{\bf 16}(13),
  pp.~9746--9752, 2008.

\bibitem{Yua08b}
Y.~Yuan, C.~Bingham, T.~Tyler, S.~Palit, T.~H. Hand, W.~J. Padilla, N.~M.
  Jokerst, and S.~A. Cummer, ``A dual-resonant terahertz metamaterial based on
  single-particle electric-field-coupled resonators,'' {\em Appl. Phys.
  Lett.}~{\bf 93}, p.~191110, 2008.

\bibitem{Bin08}
C.~M. Bingham, H.~Tao, X.~Liu, R.~D. Averitt, X.~Zhang, and W.~J. Padilla,
  ``Planar wallpaper group metamaterials for novel terahertz applications,''
  {\em Opt. Express}~{\bf 16}(23), pp.~18565--18575, 2008.

\bibitem{Hoo07}
D.-H. Kwon, D.~H. Werner, A.~V. Kildishev, and V.~M. Shalaev, ``Near-infrared
  metamaterials with dual-band negative-index characteristics,'' {\em Opt.
  Express}~{\bf 15}(4), pp.~1647--1652, 2007.

\bibitem{Wen09}
Q.-Y. Wen, H.-W. Zhang, Y.-S. Xie, Q.-H. Yang, and Y.-L. Liu, ``Dual band
  terahertz metamaterial absorber: {Design}, fabrication, and
  characterization,'' {\em Appl. Phys. Lett.}~{\bf 95}(24), p.~241111, 2009.

\bibitem{Tao10}
H.~Tao, C.~M. Bingham, D.~Pilon, K.~Fan, A.~C. Strikwerda, D.~Shrekenhamer,
  W.~J. Padilla, X.~Zhang, and R.~D. Averitt, ``A dual band terahertz
  metamaterial absorber,'' {\em J. Phys. D: Appl. Phys.}~{\bf 43}(22),
  p.~225102, 2010.

\bibitem{Liu11}
X.~Liu, T.~Tyler, T.~Starr, A.~F. Starr, N.~M. Jokerst, and W.~J. Padilla,
  ``Taming the blackbody with infrared metamaterials as selective thermal
  emitters,'' {\em Phys. Rev. Lett.}~{\bf 107}(4), p.~045901, 2011.

\bibitem{Che10}
H.~Chen, C.~T. Chan, and P.~Sheng, ``Transformation optics and metamaterials,''
  {\em Nat. Mater.}~{\bf 9}, pp.~387--396, 2010.

\bibitem{Gol11}
M.~D. Goldflam, T.~Driscoll, B.~Chapler, O.~Khatib, N.~M. Jokerst, S.~Palit,
  D.~R. Smith, B.-J. Kim, G.~Seo, H.-T. Kim, M.~D. Ventra, and D.~N. Basov,
  ``Reconfigurable gradient index using {VO}$_2$ memory metamaterials,'' {\em
  Appl. Phys. Lett.}~{\bf 99}(4), p.~044103, 2011.

\bibitem{Sch06}
D.~Schurig, J.~J. Mock, and D.~R. Smith, ``Electric-field-coupled resonators
  for negative permittivity metamaterials,'' {\em Appl. Phys. Lett.}~{\bf
  88}(4), p.~041109, 2006.

\bibitem{Pad07}
W.~J. Padilla, M.~T. Aronsson, C.~Highstrete, M.~Lee, A.~J. Taylor, and R.~D.
  Averitt, ``Electrically resonant terahertz metamaterials: {Theoretical} and
  experimental investigations,'' {\em Phys. Rev. B}~{\bf 75}(4), p.~041102,
  2007.

\bibitem{Wit10}
W.~Withayachumnankul, C.~Fumeaux, and D.~Abbott, ``Compact electric-lc
  resonators for metamaterials,'' {\em Opt. Express}~{\bf 18}(25),
  pp.~25912--25921, 2010.

\bibitem{Wit11}
W.~Withayachumnankul, C.~Fumeaux, and D.~Abbott, ``Planar array of
  electric-{LC} resonators with broadband tunability,'' {\em IEEE Antennas
  Wirel. Propag. Lett.}~{\bf 10}, pp.~577--580, 2011.

\bibitem{Pen04}
J.~B. Pendry and D.~R. Smith, ``Reversing light with negative refraction,''
  {\em Phys. Today}~{\bf 57}(6), pp.~37--43, 2004.

\bibitem{Smi04}
D.~R. Smith, J.~B. Pendry, and M.~C.~K. Wiltshire, ``Metamaterials and negative
  refractive index,'' {\em Science}~{\bf 305}(5685), pp.~788--792, 2004.

\bibitem{Oha07}
J.~F. O'Hara, E.~Smirnova, A.~K. Azad, H.-T. Chen, and A.~J. Taylor, ``Effects
  of microstructure variations on macroscopic terahertz metafilm properties,''
  {\em Act. Passive Electron. Compon.}~{\bf 2007}, p.~49691, 2007.

\bibitem{Aza08}
A.~K. Azad, A.~J. Taylor, E.~Smirnova, and J.~F. O’Hara, ``Characterization and
  analysis of terahertz metamaterials based on rectangular split-ring
  resonators,'' {\em Appl. Phys. Lett.}~{\bf 92}, p.~011119, 2008.

\end{thebibliography}
%\bibliographystyle{spiebib}

\end{document}